\documentstyle[prb,aps,graphicx]{revtex}

\DeclareMathAlphabet{\bi}{OML}{cmm}{b}{it}

\begin{document}

\draft

\twocolumn[\hsize\textwidth\columnwidth\hsize\csname@twocolumnfalse\endcsname

\title{Separation of quadrupolar and magnetic contributions to
       spin--lattice relaxation in the case of a single isotope}
\author{A. Suter, M. Mali, J. Roos, and D. Brinkmann}
\address{Physik-Institut, Universit\"{a}t
        Z\"{u}rich, CH-8057 Z\"{u}rich, Switzerland}

\maketitle
\date{\today}

\begin{abstract}
  We present a NMR pulse double--irradiation method which allows one
  to separate magnetic from quadrupolar contributions in the
  spin--lattice relaxation.  The pulse sequence fully saturates one
  transition while another is observed.  In the presence of a $|
  \Delta m | = 2$ quadrupolar contribution, the intensity of the
  observed line is altered compared to a standard spin--echo
  experiment.  We calculated analytically this intensity change for
  spins $I=1, 3/2, 5/2$, thus providing a quantitative analysis of the
  experimental results.  Since the pulse sequence we used takes care
  of the absorbed radio--frequency power, no problems due to heating
  arise.  The method is especially suited when only {\it one} NMR
  sensitive isotope is available.  Different cross--checks were
  performed to prove the reliability of the obtained results.  The
  applicability of this method is demonstrated by a study of the plane
  oxygen $^{17}$O ($I = 5/2$) in the high--temperature superconductor
  $\mathrm{YBa_{2}Cu_{4}O_{8}}$: the $^{17}$O spin--lattice relaxation
  rate consists of magnetic as well as quadrupolar contributions.
\end{abstract}

\pacs{PACS numbers: }
]


\section{Introduction}\label{intro}

The work presented in this paper has been motivated by the experience
in condensed matter nuclear magnetic resonance (NMR) experiments that
quite often both magnetic and quadrupolar time dependent interactions
are present causing spin--lattice relaxation.  The question arises
whether it is possible to deduce, directly from the experiment, the
admixture of the two different contributions to the overall
relaxation.

The literature contains mainly calculations of multiexponential
magnetization recovery laws for the case of either {\em purely}
magnetic or {\em purely} quadrupolar fluctuations, with Andrew {\em et
al.}\,\cite{andrew61} being the first to treat the case of a static
quadrupolar perturbed Zeeman Hamiltonian (spin $I=3/2, 5/2$).  These
calculations were extended to higher
spins\,\cite{tewari63,narath67,gordon78} and to the case of a static
quadrupolar
Hamiltonian\,\cite{daniel64,ainbinder78,chepin91,watanabe94}.
MacLaughlin {\em et al.}\,\cite{maclaughlin71} treated the case of a
static quadrupolar Hamiltonian ($\eta=0$) with mixed fluctuations in a
kind of perturbation expansion, whereas Rega\,\cite{rega91} presented,
for this case, an exact solution in the limit of time approaching
zero.

In a previous study \cite{suter98a}, we discussed the multiexponential
recovery for the case of a static quadrupolar perturbed Zeeman
Hamiltonian in the presence of both magnetic and quadrupolar
fluctuations under the assumption that the spin--exchange coupling can
be omitted and the eigenfunctions of the static Hamiltonian can be
approximated by Zeeman eigenfunctions.  We found that, in a
surprisingly large region of the parameter space spread by the
probabilities for magnetic and quadrupolar induced transitions, it is
almost impossible, within experimental errors, to separate magnetic
and quadrupolar contributions to the relaxation.  Instead, the
``dominant'' contribution determines the time evolution of the
recovery law, {\em i.e.} the system can approximately be described by
a {\em single} time constant, $T_{1}^{\rm eff}$.  However, it is
questionable whether this approximation is meaningful in the presence
of mixed relaxation or whether it is more appropriate to describe the
system by the {\em separate} transition probabilities.

If the nucleus under consideration has {\em two} magnetic isotopes as
in the case of copper ($^{63}$Cu and $^{65}$Cu), the admixture can be
estimated from the ratio of the relaxation times, $T_{1}$.
Nevertheless, the precision needed for a reliable interpretation of
this ratio is commonly underestimated.

In this publication we will present a method, which enables the
experimentalist to separate the different contributions of the
spin--lattice relaxation especially in the case where the element
under consideration has only {\em one} NMR sensitive isotope.  The
method involves a special initial condition of the spin system which
we call dynamic saturation and which had already been mentioned
briefly in our previous work \cite{suter98a}.

The paper is organized as follows.  The next section will introduce
the theoretical background essential to understand the method.  In
Sec.\ \ref{sec:N0} we will give the results including a discussion,
and Sec.\ \ref{sec:17O} will show experimental results of $^{17}$O NMR
in $\mathrm{YBa_{2}Cu_{4}O_{8}}$ including a more technical discussion
of the experiment.


\section{Basic relations and master equation}

For simplicity and the reader's convenience, we repeat part of our
treatment presented in Ref.\,\cite{suter98a}.  The starting point
is the following Hamiltonian:

\[
  {\cal H}_{\rm tot} = {\cal H}_{0} + {\cal H}_{1}(t),
\]

\noindent where ${\cal H}_{0} = {\cal H}_{\rm Z} + {\cal H}_{\rm Q}$
describes the time-independent (or ``static'') Hamiltonian which
comprises the Zeeman interaction, ${\cal H}_{\rm Z}$, with the
external magnetic field and the quadrupolar interaction, ${\cal
H}_{\rm Q}$, with the internal electric field gradient (EFG) tensor.
${\cal H}_{1}(t)$ takes into account fluctuations; it is the sum of a
magnetic and a quadrupolar contribution:

\begin{equation}
{\cal H}_{1}(t)   = {\cal H}_{\rm mag}(t) +
              {\cal H}_{\rm quad}(t) \label{H1},
\end{equation}
where
\begin{eqnarray}
            {\cal H}_{\rm mag}(t)  &=& -\hbar\gamma_{\rm n} \,
             {\bi{I}} \cdot {\bi{h}}(t) \nonumber \\
               {\cal H}_{\rm quad}(t) &=& \frac{eQ}{4I(2I-1)} \,
              \sum_{k=-2}^{2} V_{k}(t) T_{2k}({\bi{I}}). \nonumber
\end{eqnarray}

\noindent Here, ${\bi{I}}$ is the nuclear spin operator, ${\bi{h}}(t)$
is a fluctuating magnetic field, $V_{k}(t)$ is a component of the
fluctuating EFG, and $T_{2k}({\bi{I}})$ are spherical tensor
operators\,\cite{abragam61,slichter92}.

In Eq.\ (\ref{H1}), nuclear spin--exchange terms were omitted. If
the quadrupolar splitting, due to ${\cal H}_{\rm Q}$, is large
compared to the nuclear spin--exchange coupling, the time evolution
of the spin--lattice relaxation proceeds by the direct coupling to
the lattice. Cases where the nuclear spin--exchange terms are
important are discussed in Refs.\,\cite{andrew61,brinkmann82}.

The relaxation of the spin system towards its thermodynamic
equilibrium is described by the so--called master equation

\begin{equation}
  \frac{d}{d t}{\bi{P}}(t) = {\mathbf W}\left\{ {\bi{P}}(t) -
                               {\bi{P}}(0) \right\}.
  \label{Masterequation}
\end{equation}

\noindent Here, ${\bi{P}}(t)$ is the population vector of the
different energy levels with ${\bi{P}}(0)$ being the equilibrium
value.  The relaxation matrix, ${\mathbf W}$, is, in second order
perturbation theory, given by\,\cite{abragam61}

\begin{eqnarray}
  W_{\alpha\beta} &\stackrel{\alpha\neq\beta}{=}& \frac{1}{\hbar^{2}}
                          \int\limits_{-\infty}^{\infty}
                          d\tau \exp(i \omega_{\alpha\beta} \tau)
                          \overline{\langle \alpha | {\cal H}_{1}(\tau) |
                          \beta \rangle \langle \beta | {\cal H}_{1}(0) |
                          \alpha \rangle} \nonumber \\
        W_{\alpha\alpha} &=& - \sum_{\beta\neq\alpha} W_{\alpha\beta} ,
                          \nonumber
\end{eqnarray}

\noindent where $|\alpha\rangle$, $|\beta\rangle$ are eigenstates of
${\cal H}_{0}$ and $\omega_{\alpha\beta} = (\langle\alpha | {\cal
H}_{0} | \alpha\rangle - \langle\beta | {\cal H}_{0} |
\beta\rangle)/\hbar$ are transition frequencies.  Ensemble averages
are denoted by $\overline{\langle\ldots\rangle}$.

As long as the eigenfunctions of ${\cal H}_{0}$ can be approximated by
the eigenfunctions of a Zeeman Hamiltonian, $i.e.$ $\| {\cal H}_{\rm Z}
\| \gg \| {\cal H}_{\rm Q}\|$, the relevant relaxation matrix terms for
magnetic and quadrupolar relaxation are given as follows:

\begin{eqnarray*}
  W_{\alpha\beta}^{\rm mag} &=& J(\omega_{\alpha\beta})
             \cdot \left\{
               |\langle\alpha | I^{+} | \beta\rangle |^{2} +
               |\langle\alpha | I^{-} | \beta\rangle |^{2} \right\} \\
  W_{\alpha\beta}^{\rm quad,1} &=& J^{(1)}(\omega_{\alpha\beta})
          \cdot \left\{
      |\langle\alpha | I^{+}I_{z} +  I_{z}I^{+} | \beta\rangle |^{2}
            + \right. \\
          && \hspace{2.2cm} \left. + |\langle\alpha | I^{-}I_{z} +
                  I_{z}I^{-} | \beta\rangle |^{2} \right\} \\
  W_{\alpha\beta}^{\rm quad,2} &=& J^{(2)}(\omega_{\alpha\beta})
             \cdot \left\{
              |\langle\alpha | (I^{+})^{2} | \beta\rangle |^{2} +
               \right. \\
               && \hspace{3.3cm}  \left. +
               |\langle\alpha | (I^{-})^{2} | \beta\rangle |^{2}
             \right\}.
\end{eqnarray*}

\noindent The $J$'s are the spectral densities of the
fluctuating fields:

\begin{eqnarray*}
J(\omega) &=& \frac{\gamma_{\rm n}^2}{2} \int_{-\infty}^{\infty} d\tau
                \exp(i\omega\tau) \left[ h_{+}, h_{-} \right] \\
J^{(1,2)}(\omega) &=& \left( \frac{eQ}{\hbar} \right)^2
                \int_{-\infty}^{\infty} d\tau
                \exp(i\omega\tau) \left[ V_{+1,2}, V_{-1,2} \right]
\end{eqnarray*}

\noindent with $[A,B] = (1/2) \, \overline{(A(\tau)B(0) +
B(\tau)A(0))}$, $h_{\pm} = h_{x} \pm i h_{y}$,
$V_{\pm 1} = V_{xz} \pm
i V_{yz}$,
and $V_{\pm 2} = \frac{1}{2} (V_{xx} - V_{yy}) \pm i V_{xy}$.

If ${\cal H}_{\rm Z}$ and ${\cal H}_{\rm Q}$ are of similar magnitude,
the situation is more complicated.  The case of {\em purely magnetic
fluctuations}, for $\| {\cal H}_{\rm Z} \| \approx \| {\cal H}_{\rm Q}
\|$, has been treated by various
authors\,\cite{takigawa91b,horvatic92}.

In this paper, we will deal with the case $\| {\cal H}_{\rm Z}\|
\gg \| {\cal H}_{\rm Q}\|$ and we make the additional assumption
that the spectral densities can be approximated by a single value.
This means that the inverse of the correlation time,
$\tau_{c}^{-1}$, of the fluctuating fields is large compared to
$\omega_{\alpha\beta}$, that is $\omega_{\alpha\beta}\tau_{c} \ll
1$. One then obtains:

\begin{eqnarray}
  J(\omega) &\simeq& J(0) =: W \nonumber \\
  J^{(1,2)}(\omega) &\simeq& J^{(1,2)}(0) =: W_{1,2} \nonumber
\end{eqnarray}

\noindent and the resulting transition probabilities become

\begin{eqnarray*}
  W_{m\to m-1}^{\rm mag}    &=& W \,(I+m)(I-m+1) \\
  W_{m\to m-1}^{\rm quad,1} &=& W_{1} \,
                 \frac{(2m-1)^{2}(I-m+1)(I+m)}{2I(2I-1)^{2}} \\
  W_{m\to m-2}^{\rm quad,2} &=& W_{2} \,
                 \frac{(I+m)(I+m-1)(I-m+1)(I-m+2)}{2I(2I-1)^{2}}.
                 \label{W2}
\end{eqnarray*}

\begin{figure}[h]
        \centering
        \includegraphics[width=\linewidth]{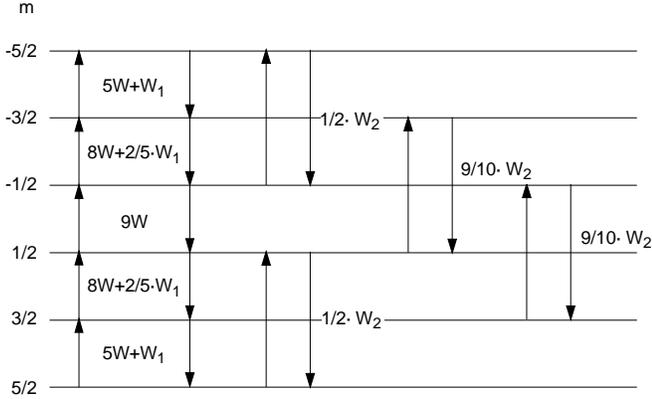}
        \caption{Transitions between the spin energy levels effected by
                 magnetic and quadrupolar spin--lattice relaxation processes
                 for $I=5/2$.}
        \label{fig:TransProb}
\end{figure}

\noindent Our calculations were performed in the high--temperature limit,
{\em i.e. $\hbar\omega_{\alpha\beta} \ll k_{B}T$}, so that a
further simplification takes place: $W_{\alpha\to\beta} \simeq
W_{\beta\to\alpha}$. Fig.\ \ref{fig:TransProb} sketches the various
transition probabilities which are possible for a 5/2 spin system.
We assume the spacings between the levels to be sufficiently
unequal to suppress spin--exchange transitions.

To solve the master equation, Eq.  (\ref{Masterequation}), it is
convenient to introduce some abbreviations.  The population of level
$m$ is $P_{m}$ and we define the difference in population between
adjacent levels by $P_{m+1/2} = P_{m+1} - P_{m}$; the equilibrium
value of this difference is $n_{0} = P_{m+1}(0) - P_{m}(0)$.  The
deviation of the population difference from its equilibrium value is
denoted by $N_{m+1/2} = P_{m+1/2} - n_{0}$; the values $N_{m+1/2}$
form the vector ${\bi{N}}$.

Given the transition probabilities as shown in Fig.
\ref{fig:TransProb}, we can write down, in compact form, the
following ``reduced'' master equation for ${\bi{N}}$:

\begin{equation}\label{ReducedMasterEquation}
  \frac{d}{d t}{\bi{N}} = {\mathbf R} {\bi{N}},
\end{equation}

\noindent where ${\mathbf R}$ is the reduced relaxation coefficient
matrix. The solution of Eq. (\ref{ReducedMasterEquation}) is of the
form

\begin{equation} \label{Njt}
  N_{j}(t) = \sum_{i} \left[\left({\mathbf E}^{T}\right)^{-1}
      {\bi{N}}(0)\right]_{i} \, E_{ij} \, \exp(t \lambda_{i}),
\end{equation}

\noindent where $\lambda_{i}$ and ${\mathbf E}$ are the eigenvalues
and the eigenvector matrix of ${\mathbf R}$, respectively, and
${\mathbf E}^{T}$ denotes the transposed of ${\mathbf E}$.
${\bi{N}}(0)$ is the vector describing the initial condition of the
spin system into which it has been brought during a certain
preparation period.

Once the $N_{j}(t)$ are known, the corresponding time dependent
magnetization, $M(t)$, is obtained:

\begin{equation} \label{Mt}
  M(t) = M(\infty) \left[ 1 - \sum_{i} a_{i}
         \exp (t \lambda_{i}) \right]
\end{equation}

\noindent and the $a_{i}$ are given by

\begin{equation}\label{ai}
  a_{i} = -\frac{1}{n_{0}} \left[ \left( {\mathbf E}^{T} \right)^{-1}
          {\bi{N}}(0) \right]_{j} E_{ji},
\end{equation}

\noindent where the index $j$ refers to the line which will be
observed, {\em e.g.} the central transition.  Usually, the irradiated
line and the observed line are the same.

We will now consider an experiment, which we call {\it dynamic
saturation}, with a special initial condition; this pulse sequence
will allow us to disentangle quadrupolar from magnetic contributions
in the spin--lattice relaxation.  We saturate a selected line, $q$,
for instance by a long comb of pulses in such a way that the comb
length, $T_{\rm tot}$, is much larger than $1/\min(W,W_{1},W_{2})$ and
that the pulse spacing, $T_{CD}$, within the comb satisfies the
condition $5 T_{2} < T_{\rm CD} \ll 1/\max(W,W_{1},W_{2})$.  This
situation contrasts with an adiabatic manipulation of the spin system
where, with the spin system initially in equilibrium, a short
radio--frequency (RF) pulse is applied to one of the transitions.  In
the case of dynamic saturation, the initial condition must be
calculated since the stimulating RF field causes transitions, with
transition rate $P_{\rm rf}$, between the levels $q+1/2$ and $q-1/2$.
Thus, for calculating the initial condition vector, the rate equation
(\ref{ReducedMasterEquation}) must be extended in the following way:

\[
  \frac{d}{d t} {\bi{N}} = \left({\mathbf R}+{\mathbf S}\right)
  {\bi{N}} + n_{0} {\bi{P}}.
\]

\noindent ${\mathbf S}$ is a square matrix with all elements zero
except $S_{q \pm 1,q} = P_{\rm rf}$, $S_{q,q} = -2P_{\rm rf}$.
${\bi{P}}$ is a vector with all the elements zero except $P_{q \pm
1} = P_{\rm rf}$, $P_{q} = -2P_{\rm rf}$.  For dynamic equilibrium,
when $d {\bi{N}}/dt = 0$, we have

\[
  {\bi{N}}(\infty) = -n_{0} \left( {\mathbf R}+{\mathbf S}
  \right)^{-1} \, {\bi{P}}.
\]

\noindent ${\bi{N}}(\infty)$, which becomes the initial condition
vector ${\bi{N}}(0)$ for solving Eq.  (\ref{ReducedMasterEquation}),
is calculated under the assumption that $P_{\rm rf} \gg
\max(W,W_{1},W_{2})$.  The exact formulas for ${\bi{N}}(0)/n_{0}$ are
given in Appendix \ref{appendix}.


\section{Transition enhancement by dynamic saturation}\label{sec:N0}

Let us deal, for the moment, with the special situation of {\it pure}
magnetic relaxation, $i.e.$ $W_{1}, W_{2} = 0$.  In this case, after
dynamic saturation, the ${\bi{N}}(0)$'s for all spin values $I \ge 1$
take the form

\[ {\bi{N}}(0)/n_{0} = [0, \ldots, 0, -1, 0, \ldots, 0]
\]

\noindent where $-1$ refers to the irradiated line.  This equation
reflects the following behavior.  On the time scale of the
spin--lattice relaxation, $T_{1} = 1/(2 W)$, the new population
differences are the same as in the case of thermodynamic equilibrium,
except for the irradiated line.  That means that a spectrum obtained
by adiabatic manipulation ($e.g.$\ in a standard spin--echo
experiment) is identical with the corresponding spectrum due to
dynamic saturation, except for the irradiated line.  This is not true
anymore in the case of mixed or pure quadrupolar relaxation as we will
show now.

The intensity of a specific transition which we observe will be
denoted as follows.  $\left.  I_{\rm ad} \right|_{m\to m-1}$ is
measured by an adiabatic pulse sequence as in the case of a standard
$\pi/2 - \pi$ spin echo experiment; $\left.  I_{\rm dyn} \right|_{m\to
m-1}^{n\to n-1}$ refers to the same transition, $m\to m-1$, however,
in the presence of dynamic saturation of the transition $n\to
n-1$. Given this notation, we define an {\it enhancement factor} by

\[ E_{m \to m-1}^{n \to n-1} =
   \frac{\left. I_{\rm dyn} \right|_{m \to m-1}^{n \to n-1}}{\left.
         I_{\rm ad} \right|_{m\to m-1}}
   = 1 + \left( \frac{{\bi{N}}(0)}{n_{0}} \right)_{m-1/2}
\]

\noindent With the results of Appendix \ref{appendix} we get, for
instance, for a spin $I=5/2$ system with the central transition being
dynamically saturated and the inner satellite being observed: $E_{\pm
3/2 \to \pm 1/2}^{1/2\to -1/2} = 1 + \mu_{5} / \zeta_{1}$.  The
enhancement factor is {\em one} in the case of pure magnetic
relaxation but it is nontrivial in the case of mixed or pure
quadrupolar relaxation.

That the enhancement factor is a nontrivial function of the relaxation
process was already noticed by Pound \cite{pound50} who used it to
show that $^{23}$Na ($I=3/2$, 100\% abundance) in $\mathrm{NaNO_{3}}$
relaxes purely quadrupolar.

\begin{figure}[h]
  \centering
  \includegraphics[width=\linewidth]{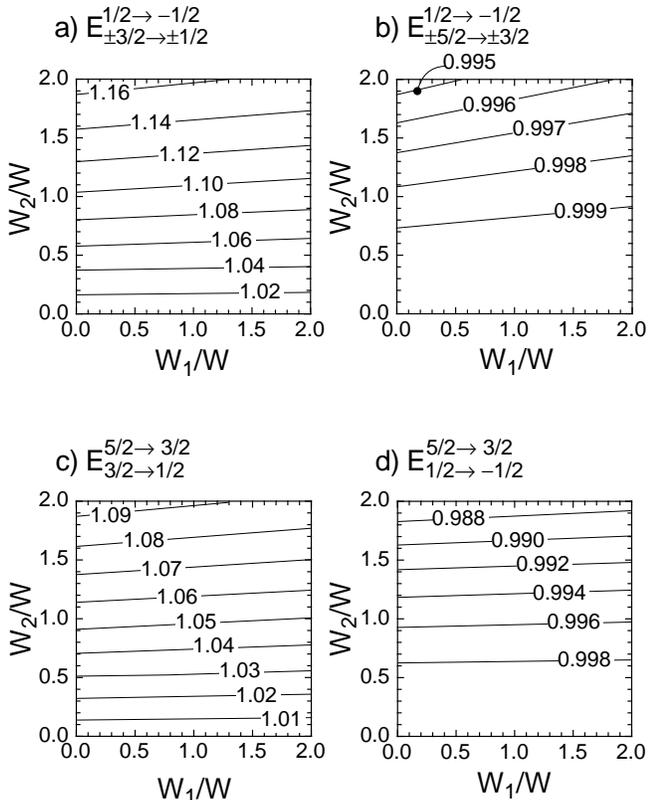}
  \caption{Contour plots of the enhancement for a spin $I = 5/2$
        system in the case of mixed relaxation.  (a) and (b) correspond to
        saturation of the central transition, (c) and (d) to saturation of
        the outer high--frequency satellite.}
        \label{fig:enhance}
\end{figure}

Fig.\ \ref{fig:enhance} shows, for a spin $I=5/2$ system with mixed
relaxation, contour plots of the enhancement factor as a function of
$W_{1}/W$ and $W_{2}/W$.  Similar results, however less pronounced,
are found for other combinations of $n\to n-1$ and $m\to m-1$.

The characteristic results are as follows. (i) The enhancement
effect is less pronounced in the case of mixed relaxation as
compared to pure quadrupolar relaxation. This makes it more
difficult to detect the effect, although not impossible because of
the very high time stability of modern spectrometers. (ii) There is
always at least one transition with an appreciable enhancement
factor [$e.g.$\ the cases (a) and (c) in Fig.\ \ref{fig:enhance}],
whereas the other transitions are ``depressed'' [cases (b) and
(d)]. This feature can be used for crosschecking the experiment if
one is able to observe both transitions at the same time; this will
be demonstrated in the next section. (iii) The enhancement factor
depends only weakly on $W_{1}/W$ since $W_{1}$ connects, except for
the $(-1/2, 1/2)$--transition, the same levels as $W$ does, which
has no effect on the enhancement function. Therefore, measuring the
enhancement yields information only about the quadrupolar $\Delta m
= 2$ transitions.

\section{Experimental details}\label{sec:17O}

We will discuss experimental details guided by our study of the
high--temperature superconductor $\mathrm{YBa_{2}Cu_{4}O_{8}}$;
these investigations will be published elsewhere \cite{suter99c}.

The experiment was performed by using a combination of two standard
pulse spectrometers together with a magnetic field of $B_{0} =
8.9945$~T, ($B_{0} \, \| \, c$). The resonant circuit was damped by
a $12\ \Omega$ resistor in order to achieve a broad frequency
range. The resonance signals were obtained by a phase alternating
add--subtract spin--echo technique similar to that one described in
Ref.\ \onlinecite{zimmermannPhD91} followed by Fourier
transformation of the spin--echo.

\begin{figure}[h]
        \centering
        \includegraphics[width=\linewidth]{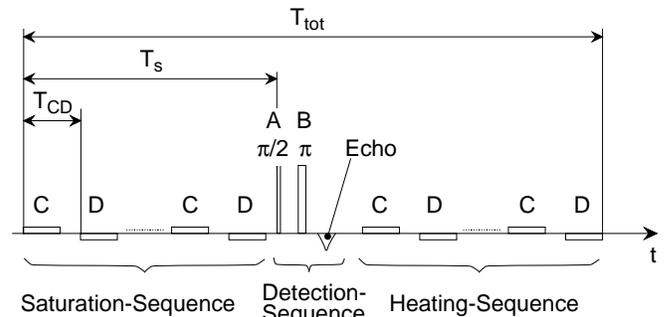}
        \caption{Pulse sequence used in the experiment. Details are
                 described in the text.}
        \label{fig:pulsseq}
\end{figure}

Each experiment consists of a certain combination of pulse
sequences which are shown in Fig.\ \ref{fig:pulsseq}. To measure
$I_{\rm dyn}$, we apply the {\it saturation} and the {\it
detection} sequence. In the first sequence, dynamic saturation of
the $n \to n-1$ transition is achieved by applying pulses $C$ and
$D$. The spacing between these pulses, $T_{CD}$, has to be much
larger than $T_{2}$ and much shorter than $T_{1}$, $i.e.$ \ $5T_{2}
< T_{CD} \ll T_{1}$. The length of these pulses is chosen very
large (20 $\mu$s) in order to saturate the $n \to n-1$ transition
only. Furthermore, we change the phase between the $C$ and $D$
pulses by $90^\circ$ to get rid of possible coherence effects. The
total length of the saturation sequence, T$_{S}$, has to be of the
order of the longest effective relaxation time. The {\it detection}
sequence is the usual spin--echo $\pi/2 - \pi$ pulse sequence.
Here, we use very intense pulses (the $\pi/2$ pulse length is about
1$\mu$s) to observe both the central transition and the
high--frequency satellites in a single shot.

In order to measure $I_{\rm ad}$, we apply, of course, the
detection sequence, but it must be supplemented by a {\it heating}
sequence. Since pulsing heats the sample, a comparison of line
intensities of different experiments (standard spin-echo and
dynamic saturation) requires constant sample temperature. This is
achieved by making the heating and the saturation sequence
identical so that the total power absorbed by the sample is the
same in either case. In other words, the total length of the pulse
sequence, $T_{\rm tot}$, is kept fixed ($T_{\rm tot} = 400$~ms in
our case). From our previous experiments on the isotope effect of
the spin gap \cite{raffa98b} we know that, at about 95~K, a
constant temperature is achieved after running the heating sequence
for about 5 minutes. A combination of all three pulse sequences is
used to cross--check our results; this will be discussed further
below.

\begin{figure}[h]
  \centering
  \includegraphics[width=\linewidth]{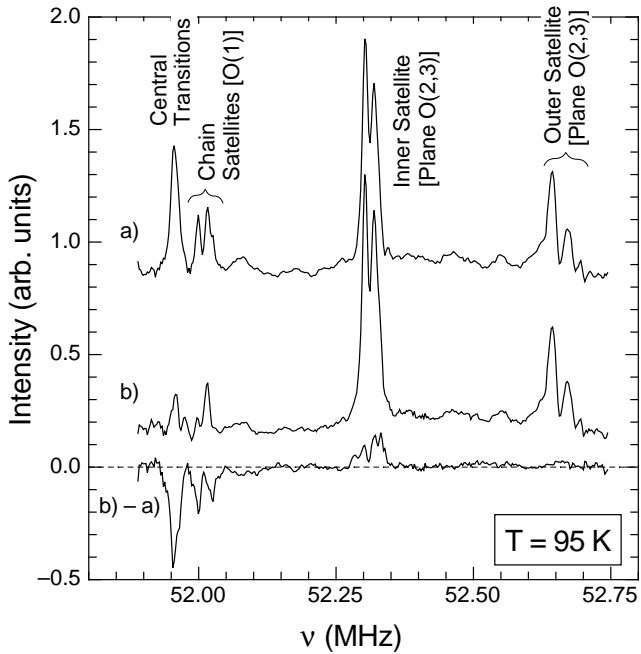}
  \caption{a) $^{17}$O central transitions and high--frequency
           satellites of plane oxygen, O(2) and O(3), and chain
           oxygen sites, O(1), in $\mathrm{YBa_{2}Cu_{4}O_{8}}$.
           b) Spectrum obtained after dynamic saturation
           of the central transition. Bottom: Difference of
           b) and a) spectrum.}
        \label{fig:diffspek}
\end{figure}

To illustrate the method, we present $^{17}$O spectra from
$\mathrm{YBa_{2}Cu_{4}O_{8}}$ taken at $T = 95$~K, see
Fig.~\ref{fig:diffspek}. This superconductor contains, beside the
apex oxygen, plane oxygen sites, O(2) and O(3), in the CuO$_{2}$
plane, where superconductivity takes place, and a chain oxygen
site, O(1). Here, we are only interested in the plane sites.
Fig.~\ref{fig:diffspek} a) shows the $^{17}$O spectrum as obtained
by Fourier transform of the standard echo. All central transitions
coincide, all O(1) satellites nearly coincide, while inner and
outer O(2,3) satellites are well separated. The splitting of these
satellites is due to the orthorhombic symmetry ($a=3.8411$
\AA, $b = 3.8718$ \AA) of the crystal.

Fig.~\ref{fig:diffspek} b) presents the $^{17}$O spectrum after
dynamic saturation of the central transition and
Fig.~\ref{fig:diffspek} c) gives the difference between the
``saturation'' and the ``standard''. Because of the short $A$ and
$B$ pulses, all transitions can be observed. Obviously, the central
transition is saturated, note the ``negative'' intensity. Whereas
the central transition of the plane oxygen is totally saturated,
the central transition as well as the satellites of chain oxygen
are not. This is due to the fact that the chain oxygen nuclei have
a considerably shorter $T_{1}$ than the plane nuclei and that the
pulse sequence was optimized for plane oxygen. That this is true
could be proven by the symmetric experiment where we dynamically
saturated the outer high--frequency satellite.

The most important feature is the remaining {\it positive}
intensity of the difference spectrum at the position of the inner
high--frequency satellite. According to the discussion above and,
in particular, the contour plot of Fig.\ \ref{fig:enhance} (a),
this intensity enhancement clearly shows that there is a
quadrupolar contribution to the relaxation of the plane oxygen
nuclei. This conclusion is supported by the fact that the intensity
at the position of the {\it outer} high--frequency satellite is
almost zero in agreement with the contour plot of Fig.\
\ref{fig:enhance} (b).

The amount of quadrupolar admixture to the overall spin--lattice
relaxation is determined as follows.  Experimentally, the intensity
enhancement of the inner satellite (when the central transition is
dynamically saturated) is 1.13(2) which then results in a ratio
$W_{2}/W = 1.4(3)$ according to Fig.\ \ref{fig:enhance} (a).  If we
saturate the {\it outer} high--frequency satellite, the intensity
enhancement of the inner satellite is 1.04(2), leading to a ratio of
$W_{2}/W = 0.7(4)$ according to Fig.\ \ref{fig:enhance} (c).  The
weighted average is: $W_{2}/W = 1.15(25)$.

It is more difficult to determine the ratio $W_{1}/W$ since the
enhancement factor $E_{\pm 3/2 \to \pm 1/2}^{1/2 \to - 1/2}$ depends
only weakly on $W_{1}$.  However, we can estimate this ratio
indirectly, to be shown below; we found that $W_{1} \leq W/3$.

\begin{figure}[h]
        \centering
        \includegraphics[width=\linewidth]{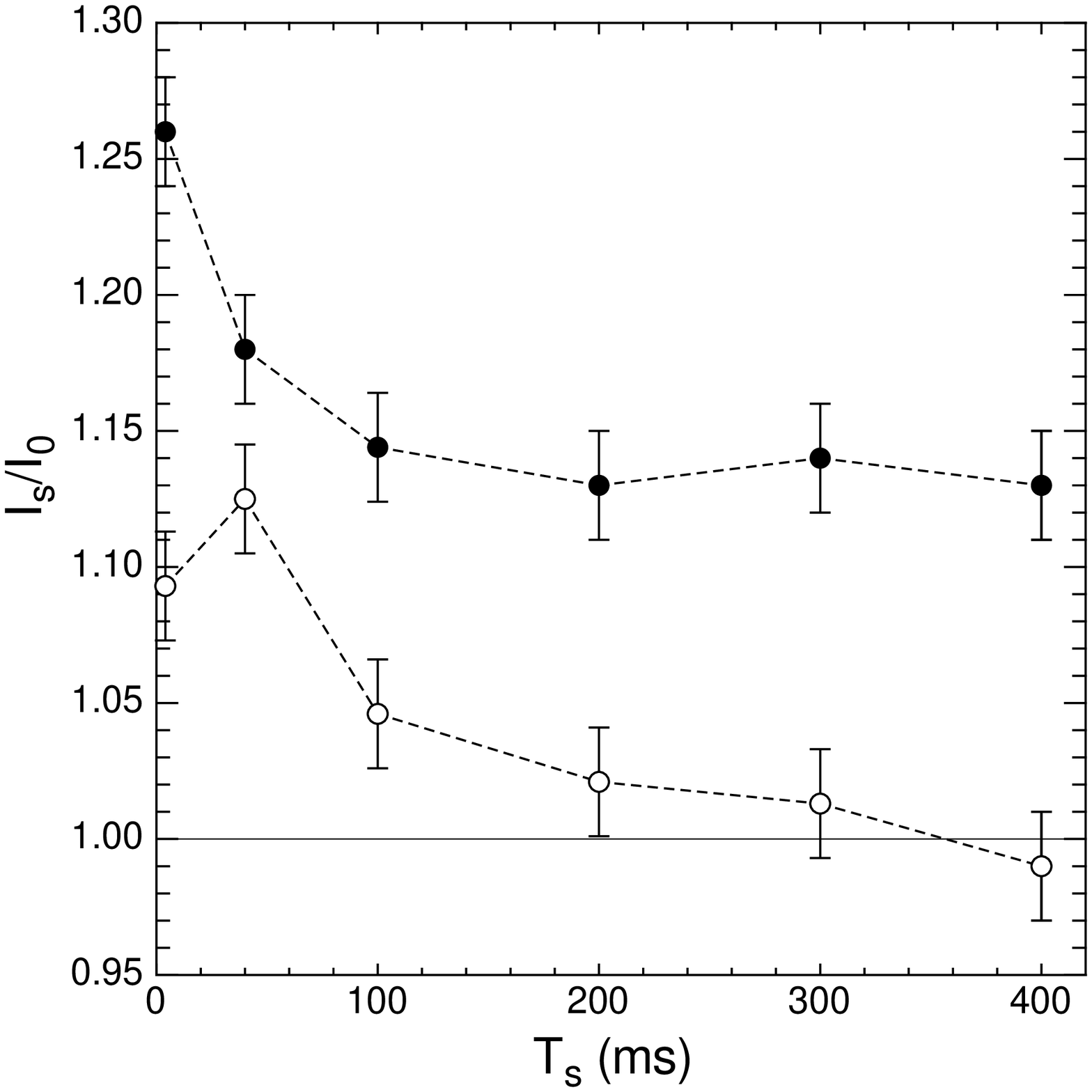} \\
        \includegraphics[width=\linewidth]{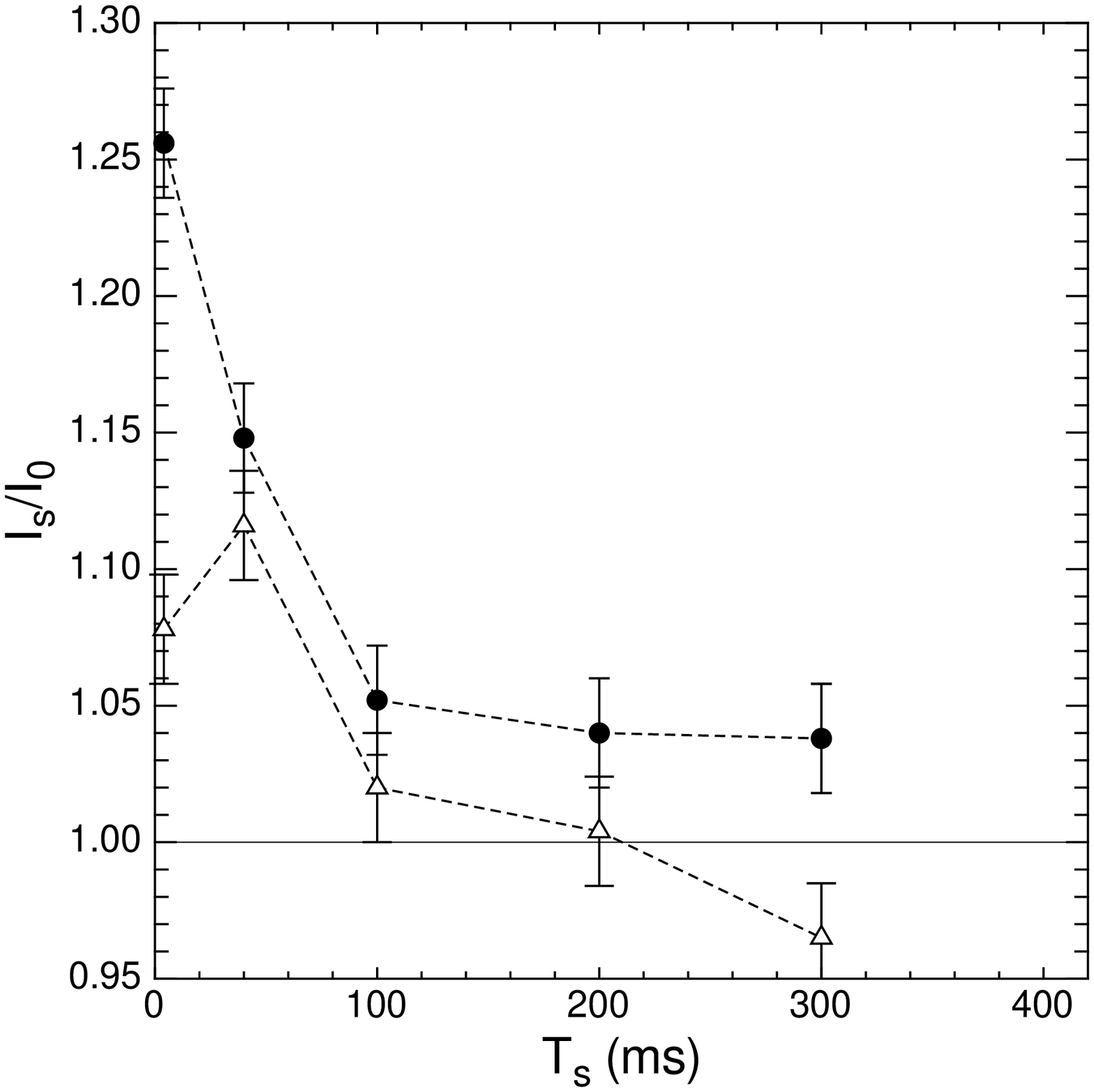}
        \caption{Top: Gradual saturation of central transition.
                 Bullets (open symbols) represent the intensity
                 enhancement of the inner (outer) high--frequency
                 satellites. Bottom: Gradual saturation of outer
                 high--frequency satellite.  Bullets (triangles)
                 represent the intensity enhancement of the inner
                 high--frequency satellite (central transition).}
         \label{fig:dynsat}

\end{figure}

To cross--check the results, we also performed an experiment with
the so--called {\em gradual saturation sequence} which involves the
application of all three pulse sequences as they are shown in Fig.\
\ref{fig:pulsseq}. $T_{s}$ is the duration of the saturation
sequence. By $I_{s}$ we denote the intensity of a line in case of
gradual saturation while $I_{0}$ is the intensity of the line in
the standard spin--echo sequence (including the heating sequence).
The {\em gradual intensity enhancement} is then defined as
$I_{s}/I_{0}$. In Fig.\ \ref{fig:dynsat}, we have plotted this
enhancement for the central transition (top figure) and for the
outer satellite (bottom) as a function of $T_{s}$. Bullets refer to
inner satellites (top and bottom), open circles to outer satellites
(top), and triangles (bottom) to the central transition.

For short $T_{s}$, the inner satellites (denoted by bullets) are
strongly enhanced as expected, since in the case of an ideal
$\pi/2$--pulse (adiabatic manipulation) at the central transition or
outer satellite, respectively, $I_{s}/I_{0}$ should reach the value
1.5. Due to spin--lattice relaxation processes this value decays
towards a limit which is (for $T_{s}$ reaching a value
corresponding to dynamical saturation) one in the case of pure
magnetic relaxation and different from one in the case of mixed or
pure quadrupolar relaxation.

The response of the outer satellite and the central transition
enhancement is retarded, since spin--lattice relaxation processes
need some time for ``pumping'' these transitions. This explains the
enhancement maximum around $T_{s} = 40$~ms, before the enhancement
starts to diminish towards one in the limit of dynamic saturation
($T_{s} \simeq 300 \ldots 350$~ms). In this limit, that is for
\mbox{$T_{s}\to 1 / \min (W, W_{1}, W_{2})$}, we have, according to
Sec.\ \ref{sec:N0} and Fig.\ \ref{fig:enhance}, $I_{s}/I_{0} \to
E_{m\to m-1}^{1/2\to -1/2}$. This is the result discussed
above.

We are now able to estimate the ratio $W_{1}/W$. $I_{s}/I_{0}$ of
the inner satellite decays very fast within the first 100~ms which
corresponds to $T_{1}^{\rm eff} = 101(5)$~ms at $T = 95$~K which we
had measured previously \cite{suter97a} by the standard inversion
recovery method. However, dynamical saturation is achieved only at
around 300\ldots 350~ms. Therefore, a slow relaxation rate must be
involved. Because $W_{2}/W \approx 1$, we conclude that this slow
relaxation rate has to be $W_{1}$, i.e.\ $W_{1} \leq W/3$. This
clearly shows that the spin--lattice relaxation process contains a
strong quadrupolar contribution which could not be detected
otherwise so far. The discussion of the origin of this effect,
which does not arise from phonons, as well as its temperature
dependence will be given elsewhere \cite{suter99c}.

\section{Summary}

We have presented a pulse double--irradiation method which allows one
to separate magnetic from quadrupolar contributions in the
spin--lattice relaxation.  The pulse sequence fully saturates one
transition while another is observed.  The clue is that the observed
transition changes its intensity if and only if a $| \Delta m | = 2$
quadrupolar contribution is present; the change is monitored with
respect to a standard spin--echo experiment.  We calculated
analytically this intensity change for spins $I=1, 3/2, 5/2$, thus
providing a quantitative analysis of the experimental results.  Since
the presented pulse sequence takes care of the absorbed
radio--frequency power, no problems due to heating arise.  The method
is especially suited when only {\it one} NMR sensitive isotope is
available.  Different cross--checks were performed to prove the
reliability of the obtained results.

The applicability of the method is demonstrated by a study of the
plane oxygen $^{17}$O ($I = 5/2$) in the high--temperature
superconductor $\mathrm{YBa_{2}Cu_{4}O_{8}}$. We showed
that the spin--lattice relaxation rate consists of magnetic as well as
quadrupolar contributions.

\acknowledgments
The partial support of this work by the Swiss National Science Foundation
is gratefully acknowledged.

\begin{appendix}
\section{Analytical formulae for initial condition vector}\label{appendix}

\subsection{Spin I=1}

\[ \frac{{\bi{N}}(0)}{n_{0}} = [\frac{\mu}{\zeta}, -1], \qquad
   \mu = 2W_{2}, \quad \zeta = 2W+W_{1}+2W_{2}
\]

\subsection{Spin I=3/2}

\noindent {\em (a) Dynamic saturation of the central transition:}

\[ \frac{{\bi{N}}(0)}{n_{0}} = [\frac{\mu}{\zeta}, -1, \frac{\mu}{\zeta}],
\qquad
   \mu = W_{2}, \qquad \zeta = 3W+W_{1}+W_{2}
\]

\noindent {\em (b) Dynamic saturation of the satellite:}

\begin{eqnarray*}
  \frac{{\bi{N}}(0)}{n_{0}} &=& [-1, \frac{\mu_{1}}{\zeta},
                              \frac{\mu_{2}}{\zeta}], \\
  \mu_{1} &=& W_{2} (3W+W_{1}+W_{2}) \\
  \mu_{2} &=& -W_{2}^{2} \\
  \zeta   &=& 12W^2+4WW_{1}+10WW_{2}+2W_{1}W_{2}+W_{2}^{2}
\end{eqnarray*}

\subsection{Spin I=5/2}

\noindent {\em (a) Dynamic saturation of the central transition:}

\[
  \frac{{\bi{N}}(0)}{n_{0}} =
[\frac{\mu_{4}}{\zeta_{1}},\frac{\mu_{5}}{\zeta_{1}},
-1,\frac{\mu_{5}}{\zeta_{1}},\frac{\mu_{4}}{\zeta_{1}}]
\]

\begin{eqnarray*}
  \mu_{4}   &=& -9W_{2}^{2}  \\
  \mu_{5}   &=& 45W_{2}(10W+2W_{1}+W_{2})  \\
  \zeta_{1} &=& 4000W^2+1000WW_{1}+40W_{1}^{2}+1100WW_{2}+ \\
            && \quad + 160W_{1}W_{2}+45W_{2}^{2}.
\end{eqnarray*}

\noindent {\em (b) Dynamic saturation of the inner satellite:}

\[
  \frac{{\bi{N}}(0)}{n_{0}} =
[\frac{\mu_{6}}{\zeta_{2}},-1,\frac{\mu_{6}}{\zeta_{2}},
                         \frac{\mu_{7}}{\zeta_{2}},\frac{\mu_{8}}{\zeta_{2}}]
\]

\begin{eqnarray*}
  \mu_{6} &=& W_{2}(8000W^3+2000W^2W_{1}+80WW_{1}^{2}+ \\
          && \quad + 3800W^2W_{2}+720WW_{1}W_{2} + 16W_{1}^{2}W_{2}+ \\
          && \quad + 440WW_{2}^{2} + 46W_{1}W_{2}^{2} + 9W_{2}^{3})  \\
  \mu_{7} &=& -9W_{2}^{2}(10W+2W_{1}+W_{2})^{2}  \\
  \mu_{8} &=& 9W_{2}^{3}(10W+2W_{1}+W_{2})  \\
  \zeta_{2} &=&
          80000W^4+36000W^3W_{1}+4800W^2W_{1}^{2}+ \\
          && \quad +160WW_{1}^{3} + 8200W^2W_{2}^{2}+46000W^3W_{2} + \\
          && \quad +16800W^2W_{1}W_{2}+
          1680WW_{1}^{2}W_{2}+32W_{1}^{3}W_{2}+ \\
          && \quad +2060WW_{1}W_{2}^{2}+
          108W_{1}^{2}W_{2}^{2}+530WW_{2}^{3}+ \\
          && \quad +64W_{1}W_{2}^{3}+9W_{2}^{4}.
\end{eqnarray*}

\noindent {\em (c) Dynamic saturation of the outer satellite:}

\[
  \frac{{\bi{N}}(0)}{n_{0}} =
   [-1,\frac{\mu_{9}}{\zeta_{3}},\frac{\mu_{10}}{\zeta_{3}},
                         \frac{\mu_{11}}{\zeta_{3}},\frac{\mu_{12}}{\zeta_{3}}]
\]

\begin{eqnarray*}
  \mu_{9} &=& W_{2}(8000W^3+2000W^2W_{1}+80WW_{1}^{2}+ \\
          && \quad + 3800W^2W_{2}+720WW_{1}W_{2} +
          16W_{1}^{2}W_{2}+ \\
          && \quad + 440WW_{2}^{2} + 46W_{1}W_{2}^{2} + 9W_{2}^{3})  \\
  \mu_{10} &=& -W_{2}^{2}(800W^2+200WW_{1}+8W_{1}^{2}+ \\
           && \quad + 220WW_{2} + 32W_{1}W_{2}+9W_{2}^{2})  \\
  \mu_{11} &=& 9W_{2}^{3}(10W+2W_{1}+W_{2})  \\
  \mu_{12} &=& -9W_{2}^{4}  \\
  \zeta_{3} &=& 128000W^4+38400W^3W_{1}+2880W^{2}W_{1}^{2}+ \\
            && \quad + 64WW_{1}^{3}+83200W^3W_{2}+
            20160W^2W_{1}W_{2}+ \\
            && \quad + 1056WW_{1}^{2}W_{2}+\frac{64}{5}W_{1}^{3}W_{2}+
            16240W^2W_{2}^{2}+ \\
            && \quad + 2744WW_{1}W_{2}^{2}+\frac{336}{5}W_{1}^{2}W_{2}^{2}
             +
            980WW_{2}^{3}+ \\
            && \quad + \frac{392}{5}W_{1}W_{2}^{3}+9W_{2}^{4}.
\end{eqnarray*}
\end{appendix}

\end{document}